\documentclass[preprint,aps,showkeys]{revtex4}
\usepackage{graphicx}
\usepackage{dcolumn}
\usepackage{bm}
\usepackage{epsfig}

\begin{document}

\title{The current-driven anomalous transports in multi-fluid
and kinetic plasma descriptions: A simulation study of anomalous transport levels}

\author{Kuang Wu Lee}%
\email{lee@mps.mpg.de}
\altaffiliation[Also at ]{Institute of Space Science, NCU Taiwan} 
\author{J\"org B\"uchner}
\affiliation{Max-Planck-Institut f\"ur Sonnensystemforschung, 
37191 Katlenburg-Lindau, Germany} 
\date{\today} 

\begin{abstract}

The generation of anomalous transport in collisionless plasma is
a significant issue in many energetic plasma environments, such as in the 
solar flares or in the strong toroidal currents of
magnetic confiment device that applies Joule heating as one of the major heating mechanisms.  
 
In most fluid models the generation mechanism and the magnetide of anomalous 
transport are usually treated as auxiliary terms external to the model description 
and are free to manipulate, 
the anomalous transport is indeed a noticeably self-generated effect exhibited in a multi-fluid system. 
The generation mechanism of current-driven anomalous transport in a multi-fluid
system is studied via a one-dimensional three-fluid simulation.
Similar to the current reduction in kinetic Vlasov simulation, this three-fluid
simulation shows that the localized electrostatic structures appear
at the nonlinear stage of plasma-wave development.  These large-amplitude structures 
appeared in both simulations play
a role as obstcles for electron bulk drifts, causing a resistive 
effect in the view that currents are reduced.  

Comparing the current relaxation levels with kinetic Vlasov
simulation of the same initial setups, it's found that there is 
a higher anomalous transport in the multi-fluid plasma, 
i.e. a stronger current reduction in the multi-fluid simulation
than in the kinetic Vlasov simulation for the same setup.  
To isolate the mechanism that causes the different anomalous transport levels, we 
hence investigated the detailed wave-particle interaction by using spectrum analysis 
of the generated waves, combined with a spatial-averaged distributions at 
different instants. 
It shows that the Landau damping in kinetic simulation takes a role that
stablizes the plasma-drifting system, when the bulk veliocity of 
electron drifts drop beneath the phase velocity of waves.
The current relaxation process stops while the relative drift velocity
between electrons is still high.

On the other hand, the current-driven anomalous transport in a multi-fluid system 
is stabilized only when the relative drifts were reduced to a very small 
value, when the system is stable in the linear dispersion analysis. 
This explains the generation of stronger anomalous transport in
a multi-fluid system than in a kinetic plasma. In the beam return-current setup used here, we found the
ability of current relaxation in multi-fluid model is around $3.5$ times
stronger than in kinetic model.  With the recognition that the Vlasov description is the more detailed
and physical description for the reality, any external dissipation term used in multi-fluid
models to stablize system should be able to bridge these differences.  
However, beside this simplified 1D electrostatic case it is necessary to examine this
effect in a 2- or 3-dimensional electromagnetic system in the view that more instabilities 
and extra degrees of freedom might change the nature of anomalous transport. 

\end{abstract}
\pacs{... }


\maketitle
\section{Introduction}
\label{sec:Introduction}
Anomalous resistivity is an influential and critical factor for many 
energetic plasma phenomena in collisionless plasma environments.
One of the examples that is influenced by Anomalous resistivity is  
the magnetic reconnection prior to the solar falres, which is a process that the topology
of the magnetic field lines in a current sheet changes and converts the 
magnetic field energy into the kinetic energy of reconnection outflow. 
Anomalous resistivity also plays an dominant role in the solar flare spectrum evolution 
of plasma distribution \cite{Lee:2008} and in the $MeV$ electron transport characteristics of the 
laser beam interaction with the ambient overdense plasma \cite{Sentoku1:2003}.

The causes of anomalous resistivity are usually studied via the comparison with the 
generation mechanisms of the classical collisional resistivity.
From the kinetic point of view the collisional electrical resistivity 
is described by classical binary collisions between charge 
particles, in which the mean-free-path is the key factor that 
determines the collisional probability within a certain time.  
Nevertheless, in most of the space plasma environments
the temporal scale of classical binary collision is usually much longer than the characteristic
period of energetic phenomena, therefore the binary collision is not sufficient to 
account for the occurrences of these transient events.  In addition, the mean-free-path of 
binary collision is usually
larger than the scale length of the spatial structures , e.g.
the collisional mean-free-path of coronal plasma is around $1 AU$ but the 
scale length of a typical solar flare loop is of the order $10-500 Mm$, which 
is much shorter than the explainable value.
Without the classical binary collisions, different species in 
plasma can exchange their energy and momentum through the 
wave-particle interaction with the generated electrostatic or
electromagnetic waves in the system. The bulk motion of charged particles
can be slowed down after the interaction with those phase space 
structures, for which the effect is identified as anomalous resistivity.

According to Boltzmann's H-theorem, collisions always push distribution
function toward the Maxwellian state to maintain a thermal equlibrium. For collisionless
plasmas, instead of binary collisions, wave-particle interactions take over the role
as a momentum converter that draws the system toward equlibrium.
In a multi-fluid system, each plasma species is described by a set of
parameters, such as drift velocity, temperature and density.  
In principle the system comprised of several species, such as a multi-fluid 
system, is deviated from equlibrium, hence a specific mechanism is expected to
draw the system back to the stable equlibrium state. 
With the bulk drift motion of charged particles, e.g. electrons in a 
current system, the distribution function is deviated from Maxwellian 
and the system is considered as free-energy supplied.

Multi-fluid plasma model has the advantage over a kinetic model on saving numerical 
resource, and more importantly a multi-fluid model conserves more informations
of wave interactions with plasma, as a fluid. Therefore
multi-fluid model has been applied to many studies of energetic plasma phenomena. 
For the plasma transport between terrestrial magnetosphere and ionosphere (M-I coupling) in the 
appearance of field-aligned current is studied via three-dimensional two-fluid ($H_{+}$, $e$)
model \cite{Gavrishchaka:1999}. Ajustable external diffusion terms ($D_{f}$) are introduced to
continuity and momentum equations to stabilize the system.  In the study of flash-like 
magnetic reconnection in astrophysical plasma, a three-dimensional multi-fluid model is 
also applied to the possible magnetic reconnection in the dusty proto-Solar nebula \cite{Lazerson:2007}.
In addition to the external-superposed collisional frequiencies between charge particles and neutral 
dust in continuity and momentum equations, anomalous resistivity is also assumed in the 
induction equation as functions of: (1) the spatial location referred to the center of current sheet, and
(2) the relative drift between ions and neutral dusts.  Plasma transport takes place in those models, even
without the external adjustable anomalous terms.  In general two collisionless transport effects are
included, the self-generated anomalous transport and the artificially-added anomalous coefficients in continuity
and momentum equations.

Although an external anomalous transport term, either assumed in a form of resistivity or collision 
frequency, the detailed informations of physical interaction between waves and particles, as well as the
weight on the above mentioned transport effects, are absent in the multi-fluid model.
An ambiguity therefore arose for dispute: with the manipulation of anomalous resistivity the simulated transport 
process do have physical correspondences or they are simply the concessions of parameter adjustments?  

Unlike the incomplete description of wave-particle interactions in multi-fluid models, the detailed 
processes of plasma evolution in the velocity space is expressed in fully kinetic model.
For turbulence and transport in large magnetic-fusion-relevant tokamaks, a systematic nonlinear 
gyrokinetic simulation study of the scalings and parameter dependences is carried on the 
ion-thermal transport rates in the work \cite{Dimits:1996}, and the geometry with strong
sheared magnetic field is considered in \cite{Dimits:1991}.
A 5D gyrokinetic simulation of plasma turbulence in a toroidal configuration with experimentally relevant parameters
is performed in \cite{Idomura:2006}.
The nonlocal heat transport by the high energy tail electrons is found to be essential for 
the preheating in laser fusion under high intensity laser irradiation \cite{Honda:1996}
Also, for the study of transport characteristics in laser fusion a three-dimensional 
Particle-In-Cell simulation describing the interaction and anomalous transport of an 
intense laser beam with a plasma slab is presented in \cite{Ruhl:2002}.


Despite these different plasma descriptions used for the anomalous transport studies, in principle
the momentum or energy transports between waves and plasma are due to the analogy collisions between
charged particles with the localized electrostatic
structures via electric or electromagnetic forces in the collisionless environments.
Current driven instabilities, such as Buneman mode or various acoustic instabilities, start 
to develop in the system.
Those localized electrostatic structures can start to grow from a small 
perturbation, like from the plasma thermal perturbation, and those electrostatic
structures can slow down the charged particles by taking their kinetic momentum
and convert into the wave energy, in the way like the classical binary collision.

With the knowledge of plasma transport in collisionless environment, anomalous resistivity can be 
determined via the momentum conversion rate.
Anomalous resistivity is a macroscopic quantity that is defined as the current reduction in collisionless plasma.
According to this definition the anomalous resistivity has been widely 
studied and discussed in fully kinetic simulations such as kinetic Vlasov 
simulation \cite{Elkina:2005,Petkaki:2006} with open boundary or via 
PIC simulation \cite{Sato:1981} on electrostatic double layer
formation in one dimensional case.  
In the fluid descriptions of plasmas, such as in resistive MHD \cite{Hesse:1996}
or even in 3D multi-fluid simulation \citep{Ganguli:2001}, the
anomalous resistivity is usually defined as a free parameter
and the generation mechanism is not concerned. 
The localized phase-space electrostatic structures are the clear exhibitions of charge separation
and particle trapping effects, and a multifluid model includes only the charge separation term 
while keeping the distribution function of each species a Maxwellian state. 

In principle a multi-fluid model can exhibit, to a certain level, the anomalous 
resistivity effect, i.e. the anomalous resistivity can be self-generated.
The external anomalous transport or dissipation terms are mainly used to stabilize the system.
However, to remove the ambiguity and dispute that what is the actual level the external anomalous should be applied
in a multi-fluid model, a detailed comparison of the generated plasma transport in fully kinetic and multi-fluid
models with the same background setup should be carried. 

In this paper we aim to study the anomalous resistivity generated by the charge separation effect in
a three-fluid plasma. Starting from the simplest electrostatic case, the nolinear evolution of current carring
system and the generated collisionless transport is examined. 
In order to compare the multi-fluid results with fully kinetic simulation, we
assumed a beam return-current system with background ions, and this setup can be an analogy to the drifting 
plasmas propagating in solar coronal loop during solar flare.  
The linear stability of the beam return-current system is presented in section \ref{DispAnaly}. 
In section \ref{kinetic_section} a kinetic Vlasov simulation is performed and the spectrum analysis
is carried out for the generated waves.  To compare the anomalous transport generated in a multi-fluid
plasma, simulation is carried for the same initial setup in section \ref{Anomalous}.  Physical 
explanation of the difference on current relaxation levels in two simulations is
provided in section \ref{Landau_damping} and a summary is given in discussion.

\begin{figure*}
\epsfig{file =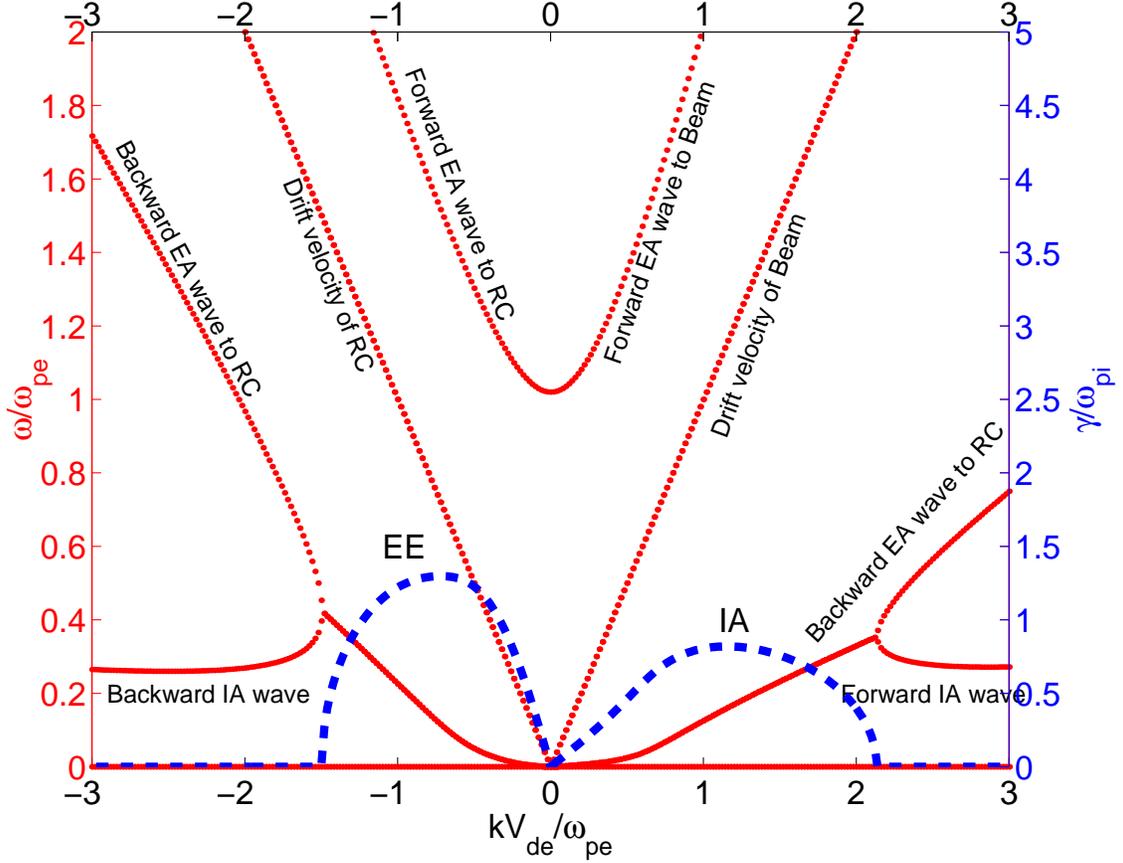, width=150mm, height=120mm}
\caption{ 
The linear multi-fluid dispersion analysis of the current-driven 
instability in a beam return current system. Two electrostatic
modes, the forward propagating ion acoustic instability and the
backward propagating electron acoustic instability, are excited in the environment.
}
\label{disp_chart}
\end{figure*}

\section{Dispersion analysis}
\label{DispAnaly}

A beam return-current system is commonly seen in many 
environments, e.g. in the solar flare X-ray loops, in the terrestrial
cusp region during magnetic substorm \cite{Velichko:1996} and in 
the laser fusion experiment \cite{Kolodner:1979}.
Usually, a beam return-current system is composed of two counter-streaming 
electron beams with one ion species stationary in the background. One of the 
electron beams which of same temperature as 
background ions is induced by the generated magnetic field, from the 
external injected beam.

In beam return-current system we assume current and charge-neutrality conditions, i.e.
\begin{eqnarray}
\sum_{\alpha} q_\alpha N_{\alpha} = 0\qquad
\sum_{\alpha} q_\alpha N_{\alpha}V_{d\alpha} = 0 
\label{conds}
\end{eqnarray} 

where $N_\alpha$ represents the densities of beams, $V_{d\alpha}$ and $v_{t\alpha}$
are the bulk velocity and thermal velocity of species $\alpha$.
In this one dimensional multi-fluid system, the electrostatic 
condition ($\vec b=(c/\omega)\vec k \times \delta \vec E=0$) 
is also assumed for the reason that large amplitude electrostatic waves 
have dominant influence on anomalous momentum 
transports in collisionless plasma. This electrostatic assumption indicates 
$\vec k\parallel \vec E$, i.e. the electric field perturbation is logitudinal
to wave vector. 

\begin{figure*}
\epsfig{file =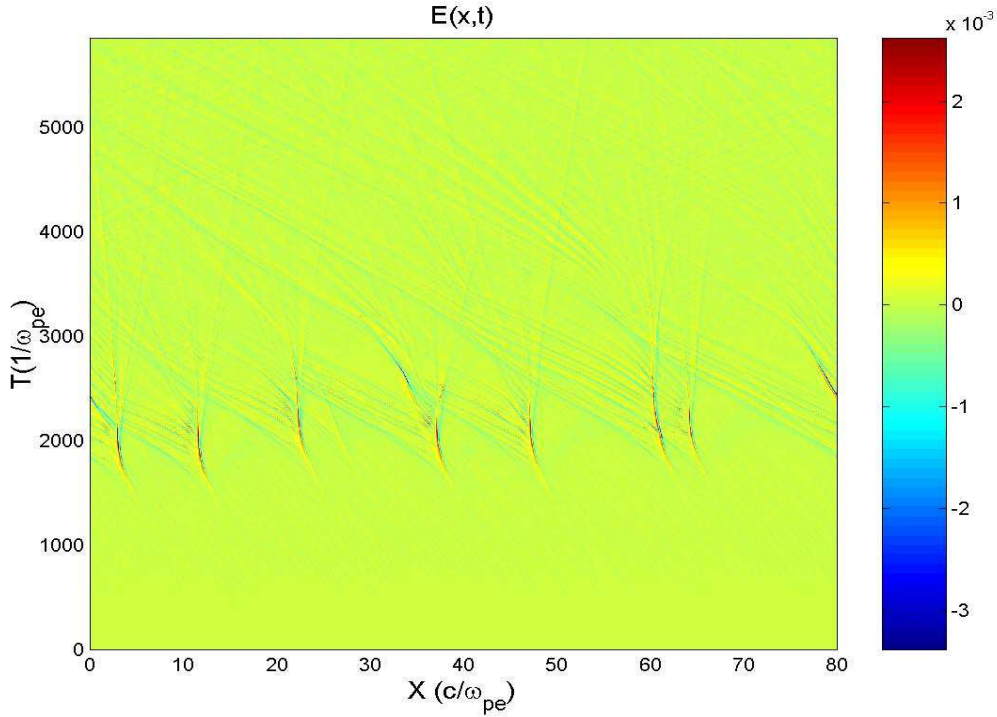, width=150mm, height=100mm}
\caption{ 
The electric field history $E_{x}(x,t)$ from fully kinetic Vlasov simulation.
At the nonlinear stage of instability development, there are several
solitary-like double layers appeared.
}
\label{XT_E_field}
\end{figure*}

The electrostatic dispersion relation for beam return-current system is:

\begin{eqnarray}
1=\frac{\omega^2_{pi}}{\omega^2} + \frac{\omega^2_{p,Beam}}
{\left(
\omega^2_{*Beam}
-k^2S^2_{Beam}
\right)}
+\frac{\omega^2_{p,RC}}
{\left(
\omega^2_{*RC}
-k^2S^2_{RC}
\right)}
\label{dispEEI}
\end{eqnarray}

The Buneman mode \citep{Buneman:1958} is a fundamental fluid-like 
instability in beam-plasma system (or ion-acoustic mode, IA, 
if thermal correction is considered).
Also, in a beam return-current system the existence of two electron populations and 
their relative drift can excite an two-stream electron-acoustic (EE) instability \citep{Gary:Book}.
 
For isothermal beam and return-current ($T_{e,Beam}/T_{e,Plasma}\approx 1$), ion-acoustic 
mode is excited along the return-current direction and an
EE two-stream instability in the beam direction.  These two electrostatic instabilities
are the two fundamental modes in the beam return-current system. The instability
dispersion analysis of the system is shown in fig.\ref{disp_chart}, and for which
the flow direction of electron beam is along the positive axis while the
return-current electron is along negative axis.
\section{Kinetic anomalous transport from Vlasov simulation}
\label{kinetic_section}
In collisionless plasma the momentum transport among species is via
wave-particle interaction, which is simply a manifestation of 
particle resonance with waves of close phase velocity or vice versa. 
In kinetic description, physics ranging from small scale particle motion 
to large scale plasma collective motion is included in the the Vlasov equation.

For the electrostatic case considered here, i.e. in the nonrelativistic zero-magnetic field 
limit, the Vlasov-Poisson equation for sepcies $\alpha$ is

\begin{eqnarray}
\frac{\partial f_{\alpha}}{\partial t}+\vec v \cdot \frac{\partial f_{\alpha}}{\partial \vec x}
+ \frac {q_{\alpha} \vec E}{m_{\alpha}}\cdot \frac{\partial f_{\alpha}}{\partial \vec v} = 0
\label{Vla_alpha}
\end{eqnarray}

The above equation neglects the binary collision term. The absence of classic collision 
indicates that, in the higher moment expressions of the Vlasov equation, the momentum and 
energy conversion between species is via wave-particle interactions.

To study the momentum and energy dissipation of beams, a
1-dimensional periodic boundary condition is applied.  The simulation
domain is set $X=[-40c/\omega_{pe},40c/\omega_{pe}]$, which is 
an intermediate length between kinetic and fluid scales.
The velocity ratios between two drifting electron and one background ion species
is as following:

\begin{eqnarray}
V_{d,Beam}=-V_{d,RC}=2V_{th,RC}
\label{parameters} \\
\nonumber 
V_{th,Beam}=2V_{th,RC}=2V_{th,ion}\\
\nonumber 
N_{RC}=N_{Beam}=N_{ion}/2
\end{eqnarray}
where $V_{d,Beam}$ and $V_{d,RC}$ are the bulk drift velocities of beam electron
and return-current electron; $V_{th,Beam}$, $V_{th,RC}$ and $V_{th,ion}$ are 
thermal velocities of beam electron, return-current electron and background ion.
The reference velocity here is $V_{th,RC}=0.01C$.  In this simulation the mass ratio
between electron and ion is chosen $m_{i}/m_{e}=25$.

This fully-kinetic simulation with the above set of parameters is run for a sufficient time
$T=5800\omega_{pe}^{-1}$. The electric field generated in this simulation is shown in a 
temporal-spatial coordinates in fig.\ref{XT_E_field}. Localized electric field structures formed and developed
at some locations. Those structures gradually developed into
large amplitude electrostatic structures that can hinder the bulk motion of electron
drifts at time $T=1600\omega_{pe}^{-1}$. As discussed in previous literature \cite{Lee:2008}
these structures are called electrostatic double layers (DLs) and they play the role
as energy converters.  The whole energy conversion process can simply catagorized at
three stages: (1) The bulk drifting motion drive the growth of local electric field perturbations
(2) The kinetic energy of bulk drifting motion is accumulated and transformed into wave energy 
(3) The localized large-amplitude electric field structures (DLs) dissipate and convert
wave energy into plasma thermal energy.

\begin{figure*}
\epsfig{file =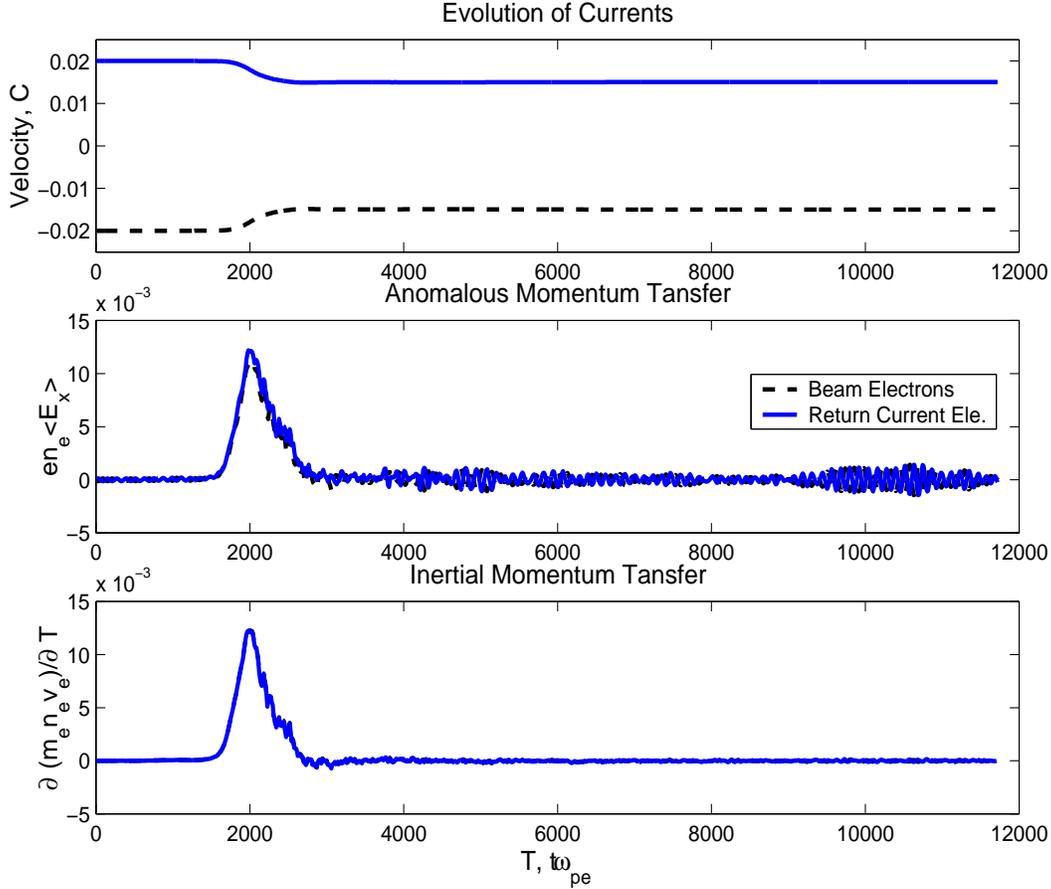, width=140mm, height=120mm}
\caption{ 
The temporal evolutions of electron bulk motions from Vlasov simulation.
The bulk velocities of electrons reduce to $3/4$ of the original values
and the dissipation of bulk kinetic energy correspond to the
appearance of electrostatic double layers, indicating that those
structures play a role as energy converters.
The second and third panels are the individual contributions of momentum
transport in this simulation, as discussed in eq.\ref{momentum_alpha_averaged_1}
}
\label{Current_history}
\end{figure*}

Current histories of beam electron and return-current electron are shown
in fig.\ref{Current_history}.  In this figure it exhibits the relaxation of electron bulk motions, at the late
stage of simulation, is about $3/4$ of the initial value.  This value, however, is
quite constant for the same velocity ratio as used in eq.\ref{parameters}.  The other 
run, which is not shown here but used a reference velocity $V_{th,RC}=0.1C$
exhibits similar relaxation level.

Take a 2nd moment on equation \ref{Vla_alpha} a fluid momentum equation for species $\alpha$ is 
obtained, which depicts the the individual contributions of momentum transfer in a collisionless
plasma. 

\begin{eqnarray}
m_{\alpha}n_{\alpha}\frac{\partial v_{\alpha}}{\partial t}+
m_{\alpha}n_{\alpha}v_{\alpha}\frac{\partial v_{\alpha}}{\partial x}
+ \frac {\partial P_{\alpha}}{\partial x} + en_{\alpha} <E_{x}> = 0
\label{momentum_alpha}
\end{eqnarray}

The first two terms indicate the momentum transport from the inertia of species $\alpha$, following
the flow velocity;
the third term $(\partial P_{\alpha})/(\partial x)$ is the momentum transport from pressure gradient and the forth 
term is the "anomalous" momentum transport from the localized electric field structures.

These individual contributions of momentum transport in a streaming plasma with
an open boundary condition are well studied in \cite{Buechner:2006}. In a simulaton domain
with an open boundary the pressure and bulk velocity of species $\alpha$, the second and third terms in 
eq.\ref{momentum_alpha} are different on two ends, therefore they contribute to the momentum transport.
In our study of comparing anomalous transports from fully-kinetic and multi-fluid plasma 
descriptions, a periodic boundary is applied
\begin{eqnarray}
\frac{\partial <m_{\alpha}n_{\alpha}v_{\alpha}>}{\partial t}+
\frac{m_{\alpha}n_{\alpha}v_{\alpha}^2|^{v_{\alpha,1}}_{v_{\alpha,2}}}{\bar{X}}+
\frac{P_{\alpha}|^{P_{\alpha,1}}_{P_{\alpha,2}}}{\bar{X}} + e<n_{\alpha}E_{x}> = 0
\label{momentum_alpha_averaged}
\end{eqnarray}

The subscript $1$ and $2$ indicate the values at the right and left boundaries.

The calculated pressures and bulk velocities on two ends are equal in a periodic boundary, hence the
second and third terms are zeros in eq.\ref{momentum_alpha_averaged}, i.e. the momentum transport 
in whole simulation domain is fully due to the anomalous term.
\begin{eqnarray}
\frac{\partial <m_{\alpha}n_{\alpha}v_{\alpha}>}{\partial t} = - e<n_{\alpha}E_{x}> 
\label{momentum_alpha_averaged_1}
\end{eqnarray}
In fig.\ref{Current_history} the momentum transport from inertial and anomalous terms are plotted in 
the second and third panels. The agreement of these two trends shows the momentum transport in this system
is primarily from wave-particle interactions.

\begin{figure*}
\epsfig{file =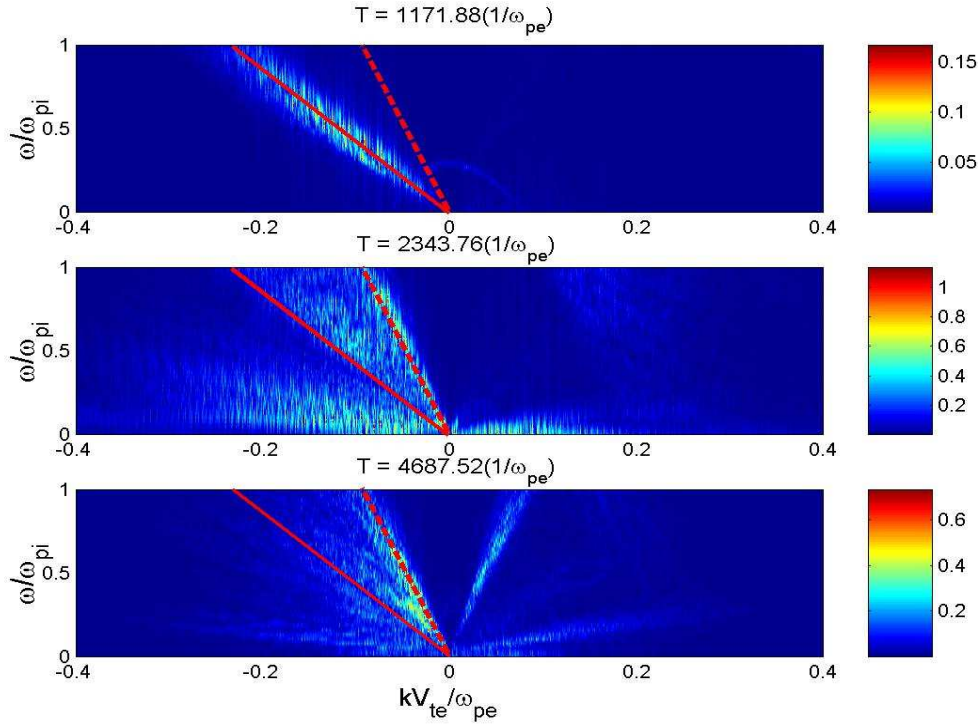, width=150mm, height=100mm}
\caption{ 
The spectrum analysis of the generated electric field perturbations
from the Vlasov simulation. The three panels correspond to wave spectra
at different instants. The phase velocity of solid lines is $\omega/k=0.01C$, the
thermal velocity of return-current electrons.  The dashed lines represent the
phase velocity $\omega/k=0.02C$, the initial drift of electron beam.
}
\label{spectrum}
\end{figure*}

With the information of localized electric field, another factor that is important 
for momentum transfer via wave-particle interaction is the spectrum of
the generated waves. To study the characteristics of waves in the system, a spectrum analysis
is applied and it demonstrates the wave intensity in the [$\omega,k$] space. 
The spectrum at different instants is shown in fig.\ref{spectrum}. It is 
clear the strongest initial electric field perturbations travel at
velocity $V_{ph}=-0.01C$ along the return-current direction. When 
the DLs appear in the instability growing phase $T=2343\omega_{pe}^{-1}$ 
the spectrum splits into very low phase velocity and high phase velocity
regions, which correspond to the velocities of DLs and fast electron holes \cite{Lee:2008}.
After the wave energy is further converted into plasma thermal energy, most of 
the electric field structures travel at the velocity of fast electron holes 
$V_{ph}=-0.02C$, indicating a post-instability stage.

\section{Self-generated anomalous transport in multi-fluid plasma and
the phase space analogy}
\label{Anomalous}

In this free-energy supplied multi-fluid plasma, system is usually stable and plasma waves 
are generated if there is small perturbation.  To mimic the thermal 
fluctuation in a warm plasma condition,
spatially random perturbation is imposed on background plasma density, and the initial 
fluctuation level is set to be $1/10^{11}$ of the background value, which is set to avoid 
the imposition of further unstable waves.  To preserve the initial charge and current 
neutrality condition, the enhancement of beam electron density corresponding a depletion
of return current electron density on the same location of simulation domain.

\begin{figure*}
\epsfig{file =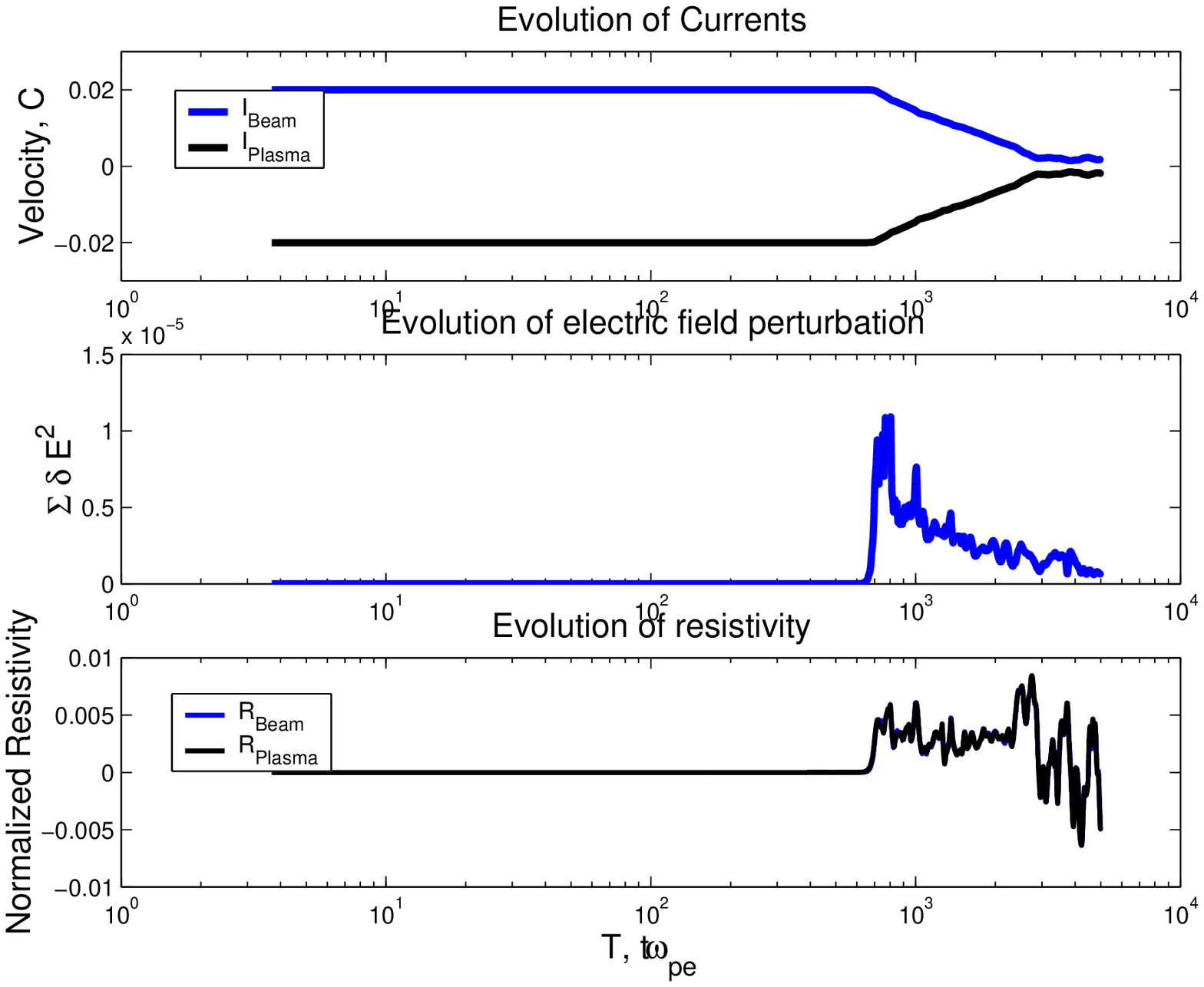, width=150mm, height=120mm}
\caption{ 
The temporal evolutions of electron bulk motion, electric field perturbation
and effective anomalous resistivities, defined as $R = (dm_{j}v_{j}/dt)/m_{j}v_{j}$.
The final currents reduced to $1/10$ of the original values, indicates a 
strong dissipation mechanism in multi-fluid plasma description.
}
\label{resistivity}
\end{figure*}

Comparing with Vlasov simulation, in order to isolate the key mechanism that causes the 
anomalous transport difference among many possible effects, the same
initial and boundary conditions are applied in the three-fluid simulation.

Similar to our understanding of anomalous transport generation in kinetic simulation, the 
plasma bulk motion is strongly influenced by the localized electric fields. Hence to study
the relationships between electric field perturbation and other physical quantities, the 
history of electron bulk motions, the temporal evolution of the electric field wave-power
and the effective anomalous transports $R_{j} = (dI_{j}/dt)/I_{j}$ 
are depicted in figure \ref{resistivity}, where $I_{j} = m_{j}n_{j}v_{j}$.
In the first panel, the current history of electron beam and return-current start to 
decrease at around $T=650 \omega_{pe}^{-1}$ and it takes about $T_{span}=2000 \omega_{pe}^{-1}$ to 
reach the lowest values.  The time of drift decreasing in Valsov simulation with same setup
is around $T_{span}=1000 \omega_{pe}^{-1}$.  The current relaxation takes place in 
the growing phase of instability and the localized electric field is 
amplified by the free energy support from bulk electron beams.
A major different feature, comparing to the kinetic results in fig.\ref{Current_history}, shown in this 
panel is the current relaxation level at the late stage of simulation.
In this multi-fluid simulation the electron drifts is eventually reduced to
around $1/10$ of the original values, which is much smaller and indicates the existence of a 
stronger current dissipation mechanism in multi-fluid plasma description. 
Presumably the discrepancy is caused, whether it is physically complete, from the model simplification of real plasma
condition in multi-fluid description. 

Because a periodic boundary condition is applied, the energy is conserved and 
transformed from one kind to another at different stages, gradually.
In the second panel of fig.\ref{resistivity} the wave energy of 
perturbative electric field is shown as a function of time. 
In line with the electric field energy evolution, a good agreement 
is shown with the development of current relaxation trend.  
This indicates the degradation of electron bulk energy corresponds to
the growth of field perturbative energy, i.e. energy is converted from bulk kinetic form
to electric field perturbation.

\begin{figure*}
\epsfig{file =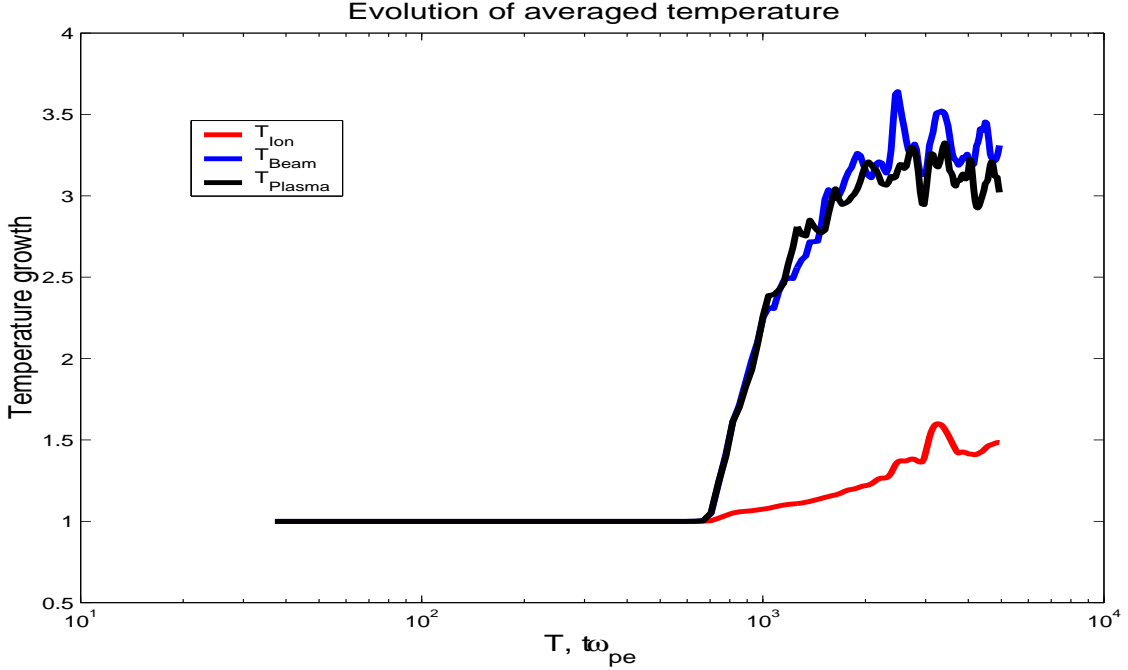, width=150mm, height=90mm}
\caption{ 
The temperature history of beam electrons, return-current electrons background and 
ions, from multi-fluid simulation.  Comparing with fig.\ref{resistivity} it shows
the temperature of both electron species increase along with the decrease
of electron bulk velocities, i.e. it indicates the bulk kinetic energy
is converted to electron thermal energy. 
}
\label{T_Beam_Plasma}
\end{figure*}

In addition, the normalized temperatures of electron beams and ions increase along with the 
relaxation of bulk drifts, too.  The temperatures of each species are shown in
fig.\ref{T_Beam_Plasma}. The energy of electron bulk motions is dissipated
and converted into electric field energy and hence the plasma thermal energy.
It is also noticed that the heating efficiency of electrons is higher than the
heating efficiency of ions. This is mainly because of the mass difference between
electrons and ions, and the frequency of generated wave is closer to the 
Doppler-shifted electron plasma frequency.

The strongest anomalous resistivity (momentum transfer rate) appears when the electric field 
energy perturbation peaks in the system, showing the dissipation of currents is the result of 
local charge separation effect, which is the most influential dragging force in multi-fluid description.  
Note that particle-trapping can not be described in this multi-fluid system, but it is self-included
in a kinetic system. The trapping effect is indeed an exhibition of Landau damping in phase 
space, i.e. those particles with close velocity as the Doppler-shifted phase velocity of 
large-amplitude plasma waves can interact and hence are trapped in the phase-space holes. 

\begin{figure*}
\epsfig{file =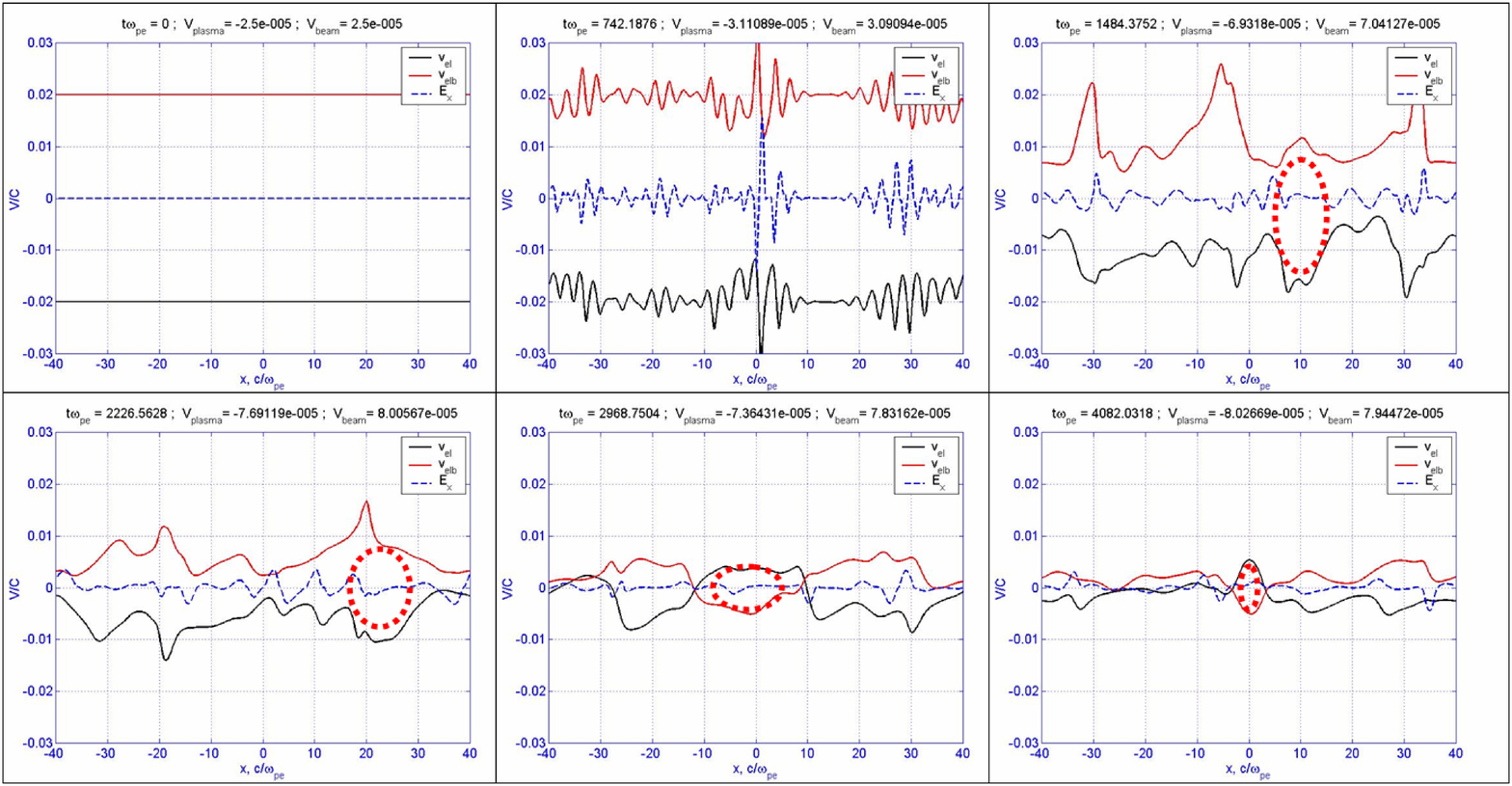, width=170mm, height=95mm}
\caption{The drift velocities of electron beams and the generated localized 
electric fields in simulation domain, at different time.  
The red-dashed circles indicate that the analogy of phase space stuctures located
at certain positions in the domain. Comparing to Vlasov simulation, the trapping effects
are excluded in multi-fluid plasma description.
}
\label{movie}
\end{figure*} 

In fluid models, plasma species are all kept maxwellian distributions and the only kinetic information 
in phase space is their temperatures.  In principle, a kinetic system can be described by infinite
constitutions of maxwellian fluids. Hence it can make an analogy of phase space structures in 
multi-fluid system. In fig.\ref{movie} the electron drifts and localized electric fields 
in the simulation domain are shown in phase-space plots, without the 
information of thermal expansion on the distributions. 
Initially the electrons drift uniformly in the domain, with an 
infinitesimal perterbations superposed onto the background. 
The initial perturbations stay small-amplitude for a period $T_{span} = 650 \omega_{pe}^{-1}$ then
an exponentially-growing phase follows. As one can see, from the left upper panel to the right 
lower panel in fig.\ref{movie}, the large-amplitude violent structures exist along with
some synchronized trends on both beam and return-current electron drifts.  In the view that
a real phase-space hole in kinetic simulation is composed of distribution functions surrounding an
empty space in the velocity space, for multi-fluid simulation the sychronized drifts can somewhat mimic 
the trend and put a circle inside certain domain, forming an analogy of phase-space structures.
Also the dissipation of individual currents is seen in a form that the mean velocities of drifts 
decrease with time.  The relaxation of bulk drifts therefore exhibits strong relation
with the localized electric field structures, which are produced by charge-separation solely 
in the electrostatic 1D multi-fluid simulation.

\begin{figure*}
\epsfig{file =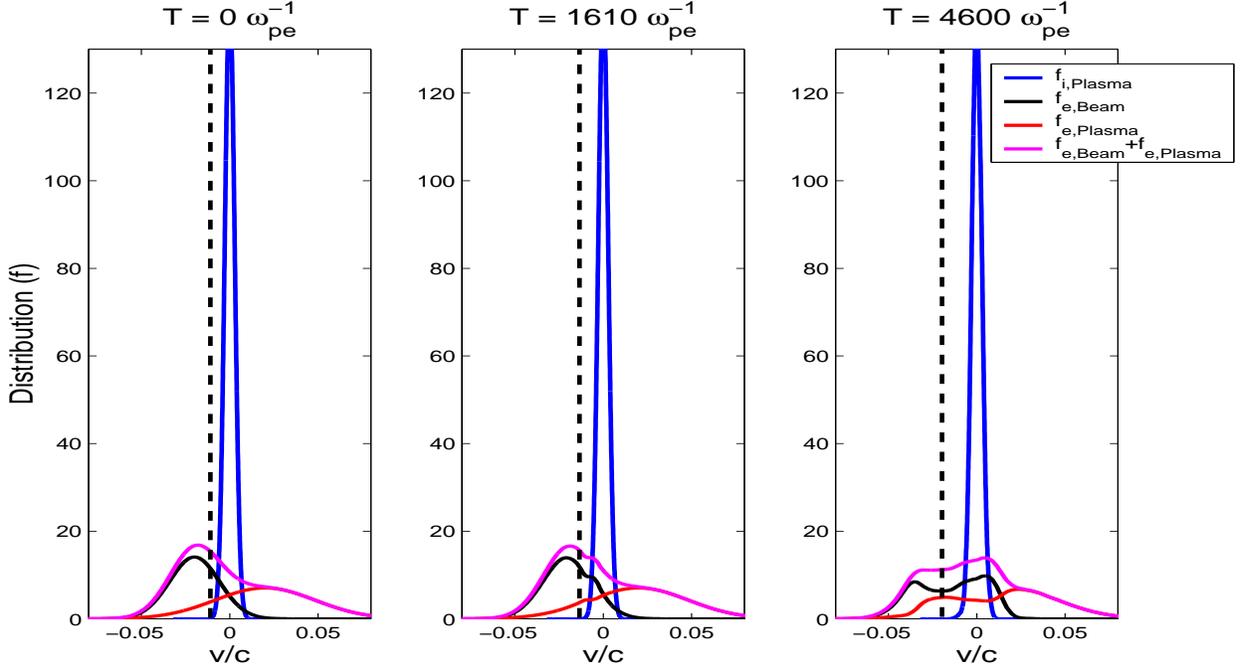, width=165mm, height=90mm}
\caption{ 
The averaged distribution functions from Vlasov simulation
at different instants.  The vertical black dashed line indicates
the phase velocity of strongest wave.  Look at the Landau damping effect 
on electrons (magenta solid line and black dashed line), it shows the system is unstable at $T = 0 \omega_{pe}^{-1}$ 
and $T=1610 \omega_{pe}^{-1}$, but the system is stabilized at $T = 4600 \omega_{pe}^{-1}$.
}
\label{Distribution}
\end{figure*}



\section{Landau damping and system stabilization}
\label{Landau_damping}

Similar to the classical Joule heating caused by collisional momentum transfer, plasma
is also heated in a collisionless system, though via a more complicated wave-particle 
interaction process instead of the binary collisions.
Nevertheless, the anomalous transports which cause the energy conversion diversify
in different plasma descriptions. In this section we would like to explore the mechanism
that results into the effective transport discrepancy in kinetic and multi-fluid plasmas.

For the same plasma setups, it shows the current relaxation is stronger in multi-fluid
description rather than in kinetic description.  Look into the primary difference of these
two regimes, the assumption of maxwellian distribution on each species distinguishes 
the multi-fluid regime from kinetic regime in the plasma evolution process.
However this assumption does not sufficiently account for the stronger dissipation 
of currents in multi-fluid plasma.  
To look at the current relaxation time spent in both
simulations, a longer relaxation process is observed in multi-fluid plasma 
before it comes to a stable state.
At the initial stage of evolution the electron drifts are high, hence the generated 
instabilities are more fluid-like in nature. Along with the development of instabilities
the bulk drifts are slowed down, therefore the instability condition becomes more 
kinetic-like because the wave phase velocity drops in the electron thermal velocity 
range, for which the Landau damping effect comes into play. The question of the generation
of different transport levels now centralize on the argument: Does the exclusion of Landau 
damping in multi-fluid regime result into the stronger current dissipation?

\begin{figure*}
\epsfig{file =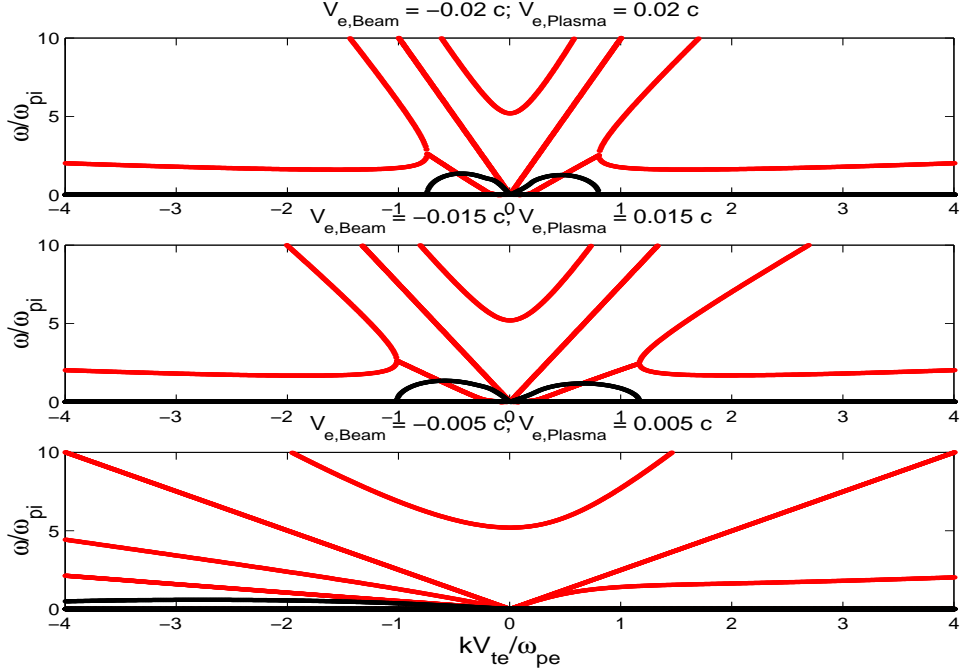, width=130mm, height=90mm}
\caption{ 
The dispersion analysis of the plasma environment at different stages
of multi-fluid simulation. In this figure, it shows the growth rates (black solid lines)
of instabilities exist even with very small drifts.  This is a result
of the exclusion of Landau correction in the large $k$ (or small wave length)
range.
}
\label{LinearAnalysis}
\end{figure*}

To verify the phypothesis, a combined spectrum analysis with an averaged distribution is used.
This method excludes the individual interaction of local plasma with wave of a wide range 
of frequencies, instead, the global interaction of plasma majority with the primary wave mode
is considered.
In section \ref{kinetic_section} the primary wave modes are deduced in the simulation spectrum 
(in fig.\ref{spectrum}).  The averaged distributions at different evolution stages, combined with 
the phase velocity of strongest wave mode, are plotted in fig.\ref{Distribution}. 
The distributions for ions, beam electrons, return-current electrons and whole electrons, are
exhibited in blue, black, red and magenta lines.
The first panel shows the distribution at $T = 0 \omega_{pe}^{-1}$, the initial plasma setup.
The initial phase velocity of wave deduced from spectrum, the Doppler-shifted electron plasma 
frequency, located in the distribution range with a nagative slope in the $f-v$ space, i.e. the
black dashed line is in the negative slope of solid magenta line. 
More particles with higher velocity interact with the wave than particles with lower 
velocity, consequently the wave is amplified and indicates the existence of an instability.

Similar situation is shown in the second panel at $T = 1610 \omega_{pe}^{-1}$, the beginning
instant of current relaxation process. As shown in fig.\ref{spectrum} the intensity of electric 
field in this stage is much stronger than in the initial.  Electrostatic DLs start to appear
at this stage as well as the fast electron holes, so there is a wider spread in the spectrum.

After the system went through a period of self-organization process, the whole electron distribution 
altered drastically (the magenta line in panel three of fig.\ref{Distribution}). Contrary to the 
stability condition at the initial stage, the particles with higher velocity are more
or equal to the amount of particles with lower velocity, indicating a Landau-stable
condition.
Because the phase velocity of waves is getting larger, or in all case at least of the value
of $|V_{ph}|\geq 0.01C$, the system becomes Landau-stable as long as the drift of electron beam 
reduced beneath this value.  Approximately the bulk drifts drop from $|V| = 0.02C$ 
to $|V| \geq 0.01C$, and this explains the current relaxation is only roughly $3/4$ of the
initial value. 

In contrast to the kinetic plasma, a multi-fluid plasma is governed by multi-fluid equations.
The Landau damping effect is not described in the multi-fluid regime, i.e. the dissipation
of current always exists as long as the instability exist.  In fig.\ref{LinearAnalysis} the
linear instability analysis is done for different drifts of electron beam, corresponding to the different 
drifts at different stages of multi-fluid simulation.  The red solid lines indicate the real 
frequencies of waves and black solid lines are the growth rates of unstable modes.  
It shows that the growth rates always exist even for a very small value of drift, indicating
the plasma waves keep growing and extracting energy from the bulk energy of electron beams.
This explains why the currents in the multi-fluid system can be reduced to a very small value.

Physically, the Landau damping effect becomes more and more significant when 
the physical process takes place in kinetic scale, i.e. large wave number $k$ or small wave length $\lambda$. 
In our case, it means when the drifts become small then growth rates shift to the large $k$ region, as shown
in fig.\ref{LinearAnalysis}.  Because the Landau damping is excluded in the linear analysis of multi-fluid
plasma, as well as in the multi-fluid simulation, the dissipation took place in kinetic scale should be carefully
treated.

\section{Discussion and conclusion}
\label{Conclusion}

The generations of anomalous transports in a 1D electrostatic kinetic Vlasov plasma and
in a 1D electrostatic three-fluid plasma are studied via numerical simulations.  
Different transport levels were observed, though the simulation setups
are the same for two different models.
With the recognition that the Vlasov description is the more detailed,
closer to physical reality description than multi-fluid model, an estimation of
anomalous transport levels in two models is given.
A combined spectrum analysis is provided in this work to explain the physical process and the 
generated difference of anomalous transport levels.

The advantage of a multi-fluid model over a fully kinetic model like given by the 
Vlasov equation is the chance to consider larger systems and complicated 
geometries, inclduing 2- and 3-dimensional ones. For a complete description and applications 
the irreversibily due to Landau damping can be introduced via an appropriate term, hence this study
provides an estimation on setting the anomalous transport intensity in multi-fluid model.

As previous literatures reported, electrostatic DLs appear in kinetic Vlasov simulation as 
energy converters that dissipate bulk kinetic energy of electron beams and heat the plasma
in the late stage of evolution.  
Spectrum analysis of the plasma waves shows the phase velocities increase from the
initial thermal velocity of return current electrons $|V_{thermal}|=0.01C$ to a value close to the initial
drift velocity $|V_{drift,RC}|=0.02C$. 

For the results from multi-fluid simulation, current relaxation and plasma heating are also 
observed, indicating that a mechanism comes into play as energy converter, and in 1D electrostatic
multi-fluid plasma it's the charge separation effects.
The major different feature shown in multi-fluid plasma, comparing to kinetic Vlasov plasma, is
the current relaxation level is much stronger than that in kinetic plasma.  From the physical model
we assumed this discrepancy of transport levels comes primarily from the neglect of 
Landau damping in the multi-fluid regime.  

To verify the assumption the Landau damping plays a major role that results the difference, we
analyzed the averaged distributions from kinetic Valsov simulation at different stages.  It 
shows the Landau damping stabilized the current relaxation when the bulk drifts drop just beneath 
the phase velocity of waves.  Because the phase velocity increases from the initial thermal velocity
of return current electrons and its value is not small, the current reduction is not severe and 
is around $3/4$ of the initial value.

For multi-fluid plasma of the same initial setup, the linear dispersion analysis as well as
the numerical scheme are described without Landau effect.  The instability and its growth rate
exist even for a very small drift, hence the waves keep extracting energy from the electron bulk
motions, causing the drastic reduction of drifts. As shown in the simulation results, the bulk drifts 
in the final stage drop to $1/10$ of the initial values. Comparing to the fully kinetic Vlasov results, the
ability of current relaxation in a multi-fluid description is around $3.5$ times stronger, for specific
velocity ratios. 

The case we considered in this work is a 1D electrostatic case, which represents the fact
that localized electrostatic structures have a dominant role in generation of anomalous 
resistivity, which has been reported in many literatures.  However, in a real multi-dimensional  
electromagnetic plasma different modes may come into play significant roles as well.  
Since the multi-fluid description of plasma has been widely considered and used in many space and
astrophysical study, to obtain a precise scenario of plasma evolution in these transient and energetic
events, a more careful and detailed study of the anomalous transport in this multi-fluid
regime should be studied and treated more carefully.

\begin{acknowledgments} 
The authors thank Dr. N. Elkina for providing the newly-developed one-dimensional
multi-fluid simulation code. K.-W. Lee acknowledges the Taiwan NSC research grant (No.95 -2911-I-008-008-2)
for supporting his stay at the MPS in Lindau. 
\end{acknowledgments}



\begin{thebibliography}{31}
\expandafter\ifx\csname natexlab\endcsname\relax\def\natexlab#1{#1}\fi


\bibitem[{B\"uchner (2006)}]{Buechner:2006}
J. B\"uchner and N. Elkina 2006, Physics of Plasmas, 13, 082304


\bibitem[{Lee et al.(2008)}]{Lee:2008}
Lee, K. W., J. B\"uchner and N. Elkina 2008, Astronomy \& Astrophys., 478, 889

\bibitem[{Sentoku1 et al. (2003)}]{Sentoku1:2003}
Y. Sentoku, K. Mima, P. Kaw and K. Nishikawa 2003, Phys. Rev. Lett., 90, 155001 

\bibitem[{Gavrishchaka et al. (1999)}]{Gavrishchaka:1999}
Valeriy V. Gavrishchaka, Supriya B. Ganguli and Parvez N. Guzdar, Journal of Geophysical 
Research 1999, 104, 22511

\bibitem[{Lazerson and Wiechen (2007)}]{Lazerson:2007}
Samuel A. Lazerson and Heinz M. Wiechen 2007, J. Plasma Physics, 74, 493-513

\bibitem[{Dimits et. al. (1996)}]{Dimits:1996}
Dimits, A. M., T. J. Williams , J. A. Byers and B. I. Cohen 1996, Phys. Rev. Lett., 77, 71

\bibitem[{Dimits et. al. (1991)}]{Dimits:1991}
Dimits, A. M., J. F. Drake , P. N. Guzdar and A. B. Hassam 1991, Phys. Fluids B, 3, 620

\bibitem[{Idomura et al. (2006)}]{Idomura:2006}
Idomura, Yasuhiro, Tomo-Hiko Watanabe and Hideo Sugama 2006, Comptes Rendus Physique, 7, 650

\bibitem[{Honda et. al. (1996)}]{Honda:1996}
Honda, M., A. Nishiguchi, K. Mima, H. Takabe, H. Azechi and S. Nakai 1996, AIPC, 369, 225

\bibitem[{H. Ruhl (2002)}]{Ruhl:2002}
Ruhl, H. 2002, Plasma Sources Science and Technology, 11, 154

\bibitem[{Lee et al.(2007)}]{Lee:2007}
Lee, K. W., J. B\"uchner and N. Elkina 2007, Physics of Plasmas, 14, 112903

\bibitem[{B\"uchner et al. (2005)}]{Elkina:2005}
J. B\"uchner and N. Elkina 2005, Space Science Reviews, 121, 237

\bibitem[{Petkaki et al. (2006)}]{Petkaki:2006}
Petkaki, P., M. P. Freeman, T. Kirk, C. E. J. Watt, and R. B. Horne 2006, J. Geophys. Res., 111, A01205

\bibitem[{Sato et al. (1981)}]{Sato:1981}
Sato, T., and H. Okuda 1981, J. Geophys. Res., 86, 3357

\bibitem[{Hesse et al. (1996)}]{Hesse:1996}
Hesse, M., J. Birn, D. N. Baker, J. A. Slavin 1996, J. Geophys. Res., 101, 10805

\bibitem[{Ganguli et al. (2001)}]{Ganguli:2001}
Ganguli, S. B., V. V. Gavrishchaka 2001, J. Geophys. Res., 106, 25601

\bibitem[{Velichko et al.(1996)}]{Velichko:1996}
V. A. Velichko, M. G. Gel'berg, and D. Yu. Zakharov 1996, Geomagnetism \& Aeronomy, 35, 5, March

\bibitem[{Paul Kolodner and Eli Yablonovitch (1979)}]{Kolodner:1979}
Paul Kolodner and Eli Yablonovitch 1979, Phys. Rev. Lett., 43, 1402

\bibitem[{Buneman(1958)}]{Buneman:1958}
Buneman, O. 1958, Phys. Rev. Lett., 1, 8

\bibitem[{Gary(1993)}]{Gary:Book}
Gary, S.~P. 1993, {Theory of Space Plasma Microinstabilities} (Theory of
  Space Plasma Microinstabilities, by S.~Peter Gary, pp.~193.~ISBN
  0521431670.~Cambridge, UK: Cambridge University Press, September 1993.)
  
\end{thebibliography}
\end{document}